\documentclass[apj]{emulateapj} 

\newcommand{\ltsim}{\,\mbox{\raisebox{0.3ex}{$<$}\hspace{-0.8em}\raisebox{-0.7ex}{$\sim$}}\,}
\newcommand{\gtsim}{\,\mbox{\raisebox{0.3ex}{$>$}\hspace{-0.8em}\raisebox{-0.7ex}{$\sim$}}\,}

\slugcomment{To appear in The Astrophysical Journal Letters}

\shorttitle{LMC disk evolution}
\shortauthors{Gallart et al.}

\begin{document}



\title{Outside-in disk evolution in the LMC.}


\author{Carme Gallart\altaffilmark{1}, Peter B. Stetson\altaffilmark{2}, Ingrid P. Meschin\altaffilmark{1},
Frederic Pont\altaffilmark{3}, Eduardo Hardy\altaffilmark{4,5} }

\altaffiltext{1}{Instituto de Astrof\'\i sica de Canarias. E-38200 La
Laguna, Spain; {\tt carme;imeschin@iac.es}}

\altaffiltext{2}{Herzberg Institute of Astrophysics, National Research
Council, Victoria, BC, Canada~V9E~2E7 {\tt Peter.Stetson@nrc.gc.ca}}

\altaffiltext{3}{Geneva University Observatory, 1290 Sauverny, Switzerland {\tt frederic.pont@obs.unige.ch}}

\altaffiltext{4}{NRAO, Chile {\tt ehardy@nrao.cl} "The National Radio Astronomy Observatory is a facility of the National Science Foundation operated under cooperative agreement by Associated Universities, Inc." }

\altaffiltext{5}{Departamento de Astronom\'\i a, Universidad de Chile, Chile}




\begin{abstract} 

From the analysis of the color-magnitude diagrams and color functions of
four wide LMC fields located from $\simeq$ 2 to 6 Kpc from the kinematic
center of the LMC we present evidence that, while the oldest population
is coeval in all fields, the age of the youngest component of the dominant 
stellar population  gradually increases with galactocentric distance,
from currently active star formation in a field at 2\degr3, to 100 Myr, 0.8 Gyr,
and 1.5 Gyr in fields at 4\degr0, 5\degr5, and 7\degr1, respectively. This outside-in
quenching of the star formation in the LMC disk is correlated with the
decreasing HI column density (which is $\le 2\times 10^{20} {\rm
cm}^{-2}$ in the two outermost fields with little or no current star
formation). 
Other work in the literature hints at similar behavior in the stellar 
populations of irregular galaxies, and in M33. This is observational 
evidence against the inside-out disk formation scenario in low-mass
spirals and irregular galaxies. Alternatively, it could be that the age 
distribution with radius results from interplay between the 
evolution with time of the star-forming area of the LMC and the subsequent 
outward migration of the stars. 

\end{abstract}


\keywords{galaxies: formation and evolution; galaxies: individual (LMC); Magellanic Clouds}


\section{Introduction} \label{intro}

Modern hydrodynamical simulations suggest that, within the LCDM
cosmology,  galaxies form from the inside out, i.e., their inner parts formed
first, and they subsequently grew up to their present-day size (e.g. Brook
et al.\ 2006; Ro\u{s}kar et al.\ 2008). This can be interpreted either in
terms of the age of the oldest population, which would be younger toward
the external parts, or in terms of the mixture of populations, which would
be younger on average there. However, the formation of a realistic disc
galaxy within such simulations remains elusive (e.g. Governato et al.
2007; Abadi et al. 2003) and it is essential to augment such studies with
detailed observations. 


Observationally, two main avenues have been used to investigate
disk formation and evolution. The first relies on the direct
comparison of the actual physical sizes (or related parameters) of high-z
disk galaxies with those of nearby galaxies (e.g. Trujillo \& Aguerri 
2004; P\'erez 2004; Reshetnikov et al. 2003). The second focuses  on
nearby galaxies, and  typically uses indicators of the ratio of recent to
total star formation rate (SFR) as a function of radius (e.g.,
Mu\~noz-Mateos et al.\ 2007; Taylor et al.\ 2005). Taylor et al.\ (2005)
found, in agreement with previous studies, that late-type spirals and
irregular galaxies become on average redder outward. Since no plausible
galaxy formation theory predicts positive metallicity gradients, they
offer stellar age effects or dust as possible causes of this effect. The
use of integrated properties makes it difficult to derive mass fractions
as a function of age (and thus to distinguish between these two
possibilities), or to determine the age of the oldest population as a
function of radius. A direct census of stellar ages in a resolved galaxy
would provide the necessary information to trace the
characteristics of the stellar population as a function of radius, and
thus to provide direct insight on formation and evolution mechanisms.

Nearby galaxies offer this possibility. Indeed, if the oldest main
sequence (MS)  turnoffs are reached, the age of the oldest population as a
function of radius can be measured and a quantitative age profile can be
determined at each radius. In dIrr galaxies, a young  population is
routinely found in their central parts, but it disappears  in the
outskirts (e.g. Bernard et al.\ 2007; Vansevicius et al.\ 2004). In the
cases where a CMD reaching the oldest MS turnoffs is available, it has
been found that an intermediate age population extends to large distances
from the center, and that it very gradually disappears toward the outer
part (S. Hidalgo et al.\ 2008, in preparation; Gallart et al.\ 2004;
No\"el \& Gallart 2007), suggesting a "shrinking scenario" of star
formation in dwarf galaxies (Hidalgo et al. 2003). A similar result is
found in the case of M33 (see Barker et al.\ 2007). 

In this paper we analyze four 35\arcmin $\times$
35\arcmin \ fields in the LMC at different galactocentric distances, using
CMDs reaching the oldest MS turnoffs, in order to shed light on its
stellar population gradients and thus on its formation and evolution.

\section{Observations and data reduction} \label{obs}

We obtained V and I images of four LMC fields with the Mosaic II camera on the
CTIO Blanco 4m telescope in December 1999 and January 2001.  Fields
were chosen to span a range of galactocentric distances, from
$\simeq$ 2.3\degr\ to 7.1\degr\ (2.0 to 6.2 Kpc) northward from the kinematic center
of the LMC. 
Note that the position angle of the LMC bar is  120\degr\ , and its center is
slightly offset to the southeast of the  kinematic center. We will name
the fields according to their RA and DEC (J2000.0) as
LMC0512-6648, LMC0514-6503, LMC0513-6333 and LMC0513-6159, in order of
increasing galactocentric distance.


The Mosaic frames were reduced in a standard way, using the MSCRED package
within IRAF\footnote{IRAF is distributed by the NOAO, which is operated by
the AURA, Inc., under cooperative agreement with the NSF.}.
Profile-fitting photometry was obtained with the DAOPHOT/ALLFRAME suite
of codes (Stetson 1994) and calibrated to the standard system using observations
of several Landolt (1992) fields obtained in the same runs. Finally,
a large number of artificial star tests were performed in each frame
following the procedure described in Gallart et al.\ (1999); these are used
both to derive completeness factors and to model photometric
errors in the synthetic CMDs. I. Meschin et al.\ (2008, in preparation)
will provide a detailed description of the observations and data
reduction.

\section{Stellar population gradients in the LMC} \label{gradients}

\begin{figure}   
\plotone{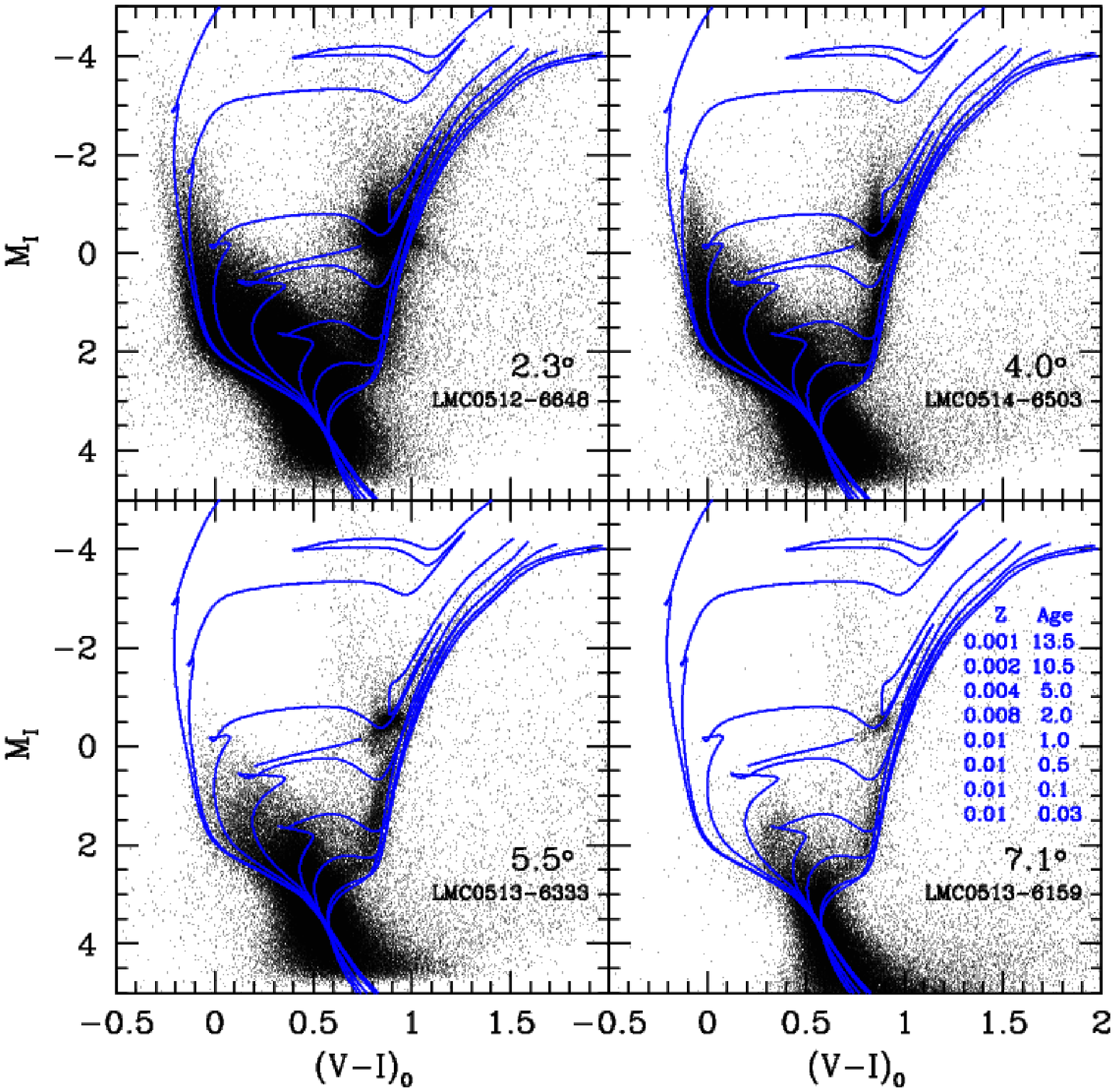} 
\figcaption{$[(V-I)_0, M_I]$ CMDs for the four fields. Isochrones with
ages and metallicities as labelled, and a zero-age horizontal-branch of
Z=0.001 by Pietrinferni et al.\ (2004) have been superimposed. A distance
modulus of $(m-M)_0=18.5$ and E(B-V)=0.10, 0.05, 0.037 and 0.026
magnitudes, respectively, have been assumed. For the two outermost fields,
the reddening values given by Schlegel et al. (1998) have
been used. For the two innermost fields their value is not accurate and we
have estimated a mean value of the reddening by requiring a good fit of
the CMD by the same isochrones. The tightness of the sequences, and in 
particular of the RC, in the different CMDs, indicate that differential 
reddening in these fields is small, except for the innermost field, where 
the whole CMD gets noticeably blurred, and a tail of reddened stars is 
clearly visible in the RC area.  
\label{cuatroiso}}   
\end{figure}

Figure~\ref{cuatroiso} shows the $[(V-I)_0,M_I]$ color-magnitude diagrams
(CMD) of the four LMC fields, with isochrones superimposed. The number of
stars observed, with good quality photometry, down to $M_I \lesssim 4$ in
each field,  in order of increasing galactocentric distance are 300000, 
214000, 86000 and 39000 respectively. All the CMDs reach the oldest MS
turnoff ($M_I\simeq$3.0) with good photometric accuracy, and completeness
fractions over 75\% (except for the innermost field, in which crowding is
very severe).  The two innermost fields show a CMD with a prominent, bright MS
and a well populated red clump (RC) typical of a population which has had
ongoing star formation from $\simeq$13 Gyr ago to the present time. The
two outermost fields clearly show a fainter MS
termination, indicative of a truncated or sharply decreasing star
formation in the last few hundred Myr or few Gyr, respectively (see
below). No extended horizontal branch is observed in any of the fields but 
all fields host a number of stars redder than the RGB tip (and redder than
the color interval shown in the figure), which are candidate AGB stars. 


\subsection{The comparison with isochrones}

In Figure~\ref{cuatroiso}, the observed CMD for each field is compared
with isochrones from the overshooting set of the BaSTI library
(Pietrinferni et al.\ 2004)\footnote{The new 2008 version of the BaSTI
stellar evolution models (see http://www.oa-teramo.inaf.it/BASTI) is used
through the paper. This set shows a much better agreement with other
stellar evolution models than the older one.}.  The metallicities of
the isochrones have been chosen to approximately reproduce the common
chemical enrichment law for the same fields derived by Carrera et al.\
(2007), using CaII triplet spectroscopy. Note how well this combination of
ages and Z reproduce the position and shape of the RGB.

The area around the old MS turnoff is well populated in all four CMDs,
indicating that star formation started at about the same time in all
four fields, or that old stars have been able to migrate out
to the galactocentric radii observed here. The 13.5 and 10.5 Gyr
isochrones of Z=0.001 and Z=0.002, respectively, almost overlap due to the
substantial metallicity increase measured in this period. The main stellar
population differences among the fields are instead related to the recent star
formation history. The MS of the innermost field, LMC0512-6648 is well
populated  up to the 30 Myr isochrone or younger. Field LMC0514-6503 shows
a MS populated up to the same isochrone,  but there is an apparent change
in the density of stars at the position of the $\simeq$ 100 Myr old
isochrone that might indicate a lower SFR from 100 Myr ago
to the present time. In the next field, LMC0513-6333, an apparent change
in the stellar density on the MS occurs at age $\simeq$ 0.8 Gyr while in
the outermost field, LMC0513-6159, it occurs at age $\simeq$ 1.2 Gyr. The
comparison with isochrones, therefore, indicates a gradual increase in the
age of the youngest component of the bulk of the stellar population with
galactocentric radius. No obvious gaps indicating a discontinuous star
formation history are present in the MS of any of the fields.

\subsection{CFs}

\begin{figure}   
\plotone{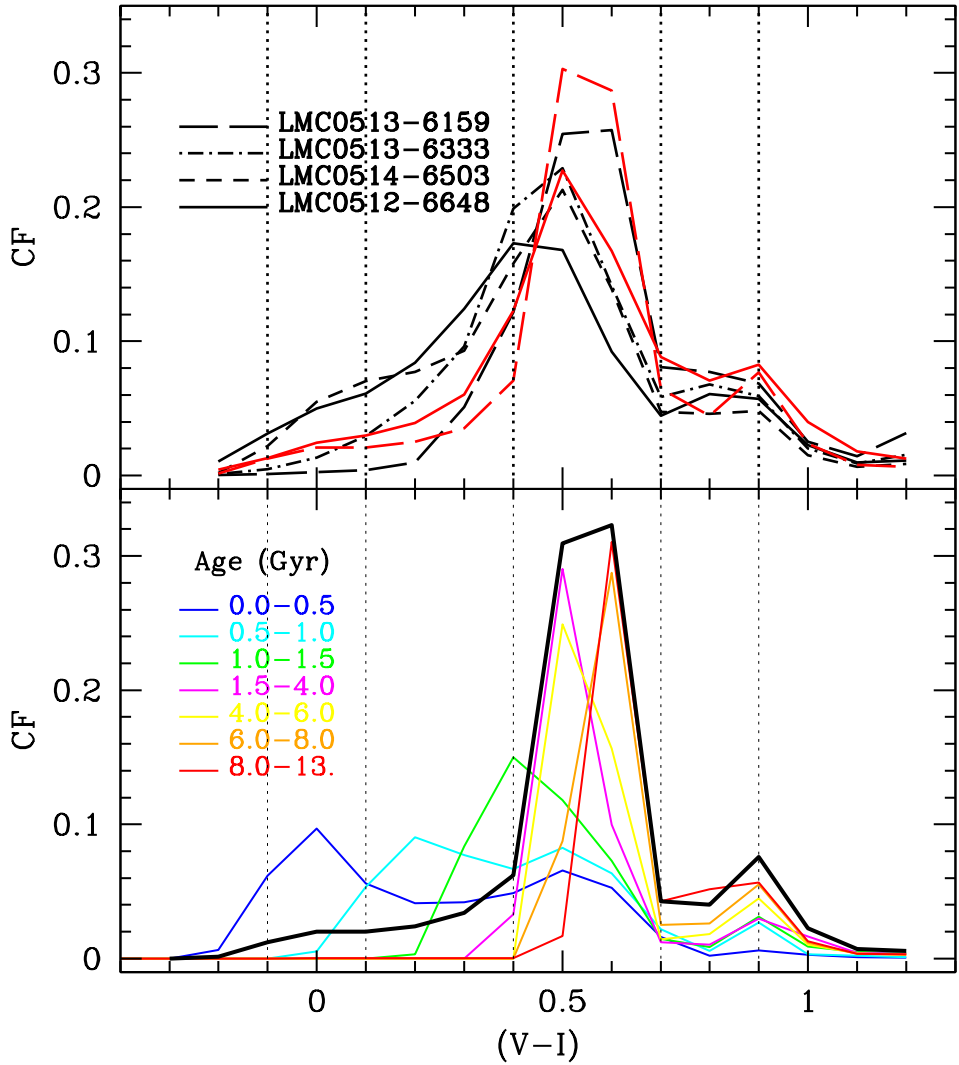}    
\figcaption{{\bf Lower panel.} Black line: CF of a synthetic CMD
computed assuming the Z(t) derived by Carrera  et al.\ (2007) using the Ca
II triplet, a constant  SFR from 13 Gyr ago to the present
time, a Kroupa (2001)  IMF and 50\% of binaries ($q\ge0.7$). The BaSTI
stellar evolution models  (Pietrinferni et al.\ 2004) have been assumed and
the CF has  been normalized to unity. Color lines: CF of the same
synthetic  CMD within limited age ranges, as labelled. Here, the
normalization factor  of each CF is half that of the CF for the whole 
range of ages. {\bf Upper panel.} Black lines with different line types:
observed CFs for each LMC field. Red lines: CF of the global synthetic CMD
used in the lower panel, but with observational errors simulated using
completeness information from the outermost (long dashed lines), and the 
innermost (solid line) fields, which are the lower and upper limit for the 
observational errors, respectively. In all cases, the CF integrates the 
information in the CMD between $M_I=-4$ and $M_I=3.5$. Vertical dotted lines 
have been drawn in order to guide the eye in the comparison between the 
synthetic and the  observed CFs.   
\label{colorfunc}}      
\end{figure}

The use of the color function (CF) for a semi-quantitative assessment of
the stellar population present in a CMD has been discussed in detail by
Gallart et al\ (2005) and No\"el et al.\ (2007). The CF accounts naturally
for the fact that the stellar density is different in the various fields.
The top panel of Figure~\ref{colorfunc} represents the observed CFs of
the four LMC fields. Red lines represent the CF of a synthetic CMD
computed using IAC-STAR (Aparicio \& Gallart 2004), assuming a constant
SFR from 13 Gyr ago to the present time and the Z(t) derived by Carrera et
al.\ (2007). Observational errors have been simulated based on the
artificial star tests performed on the outermost and innermost fields. As
in the case of the outermost field, the errors corresponding to the
intermediate fields do not substantially change the overall shape of the
CF.  The bottom panel displays the CF for the same synthetic CMD without
any errors simulated (black thick line), together with the CF
corresponding to stellar populations within limited age ranges, as
labelled (see the Figure caption for more details). 

As discussed in No\"el et al.\ (2007), the CF of a synthetic CMD computed
assuming a constant SFR can be divided into three main features: a blue
elevation composed by young stars, a main central peak which contains
stars of all ages, and a red peak which corresponds to the RGB and RC.
Since metallicity affects the RGB and RC position more strongly than 
the other CMD features, the coincidence in color between the
observed and the synthetic red peaks indicates that the assumed chemical
enrichment law is consistent with the data\footnote{The fact that
the observed CFs reach higher values in the color range (V-I)=0.7--09 than
the synthetic CF may be attributed to the foreground contamination, which
is relatively important in this color range.}. Once this metallicity
agreement is verified, the shape of the CF mainly provides information on
age and relative importance of the stellar populations younger than 
$\simeq 1.5$ Gyr.

Note that redder than $(V-I)\simeq 0.3$ the position and width of the
LMC0513-6159  CF are very similar to those of the synthetic CMD; bluer
than this color  the observed CF doesn't show the blue elevation composed
of young stars. This indicates a significant decrease in star formation
around $\simeq$ 1.5 Gyr ago, in agreement with the information provided by
the isochrones. For the remaining fields, the main peak blue side is
shifted to the blue with respect to that of the synthetic CF, possibly
indicating enhanced star formation, as compared to a constant SFR, in the
age range $\simeq$ 4--1 Gyr. The blue elevation of field LMC0513-6333 is
lower than the synthetic one for $(V-I)\ltsim 0$, confirming reduced star
formation in this field starting $\gtsim 0.5\,$Gyr ago. Finally, the CFs
of fields LMC0514-6503 and LMC0512-6648 show a blue elevation
substantially enhanced compared to a constant SFR, indicating increased
star formation  from $\simeq$ 1--1.5 Gyr ago to the present time. 

The CFs, therefore, confirm the information provided by the isochrones:
(a)~a possible truncation of the bulk of the star formation  in the
outermost fields LMC0513-6159 and LMC0513-6333 at ages $\simeq$ 1.5 and
0.8 Gyr respectively; and (b)~the suggestion of enhanced star formation
(as compared to a constant SFR) in the innermost fields LMC0514-6503 and
LMC0512-6648 in the same time range, likely having started even earlier,
$\sim$ 4 Gyr ago. Enhanced star formation starting around 4 Gyr ago
possibly extended also to field LMC0513-6333; however in this case it
would have been followed by the truncation discussed above.

\section {Discussion}

We have presented CMDs and CFs of four wide LMC fields spanning
galactocentric distances, from $\simeq$ 2 to 6 Kpc northward from the
center of the LMC. The oldest population is of similar age in all fields
and the CFs of the three innermost fields indicate a likely enhancement of
the SFR $\simeq$ 4 Gyr ago, in agreement with former studies (e.g.,
Carrera et al.\ 2007; Holtzman et al.\ 1999; Olsen 1999; Bertelli et al.\
1992). Finally--- and most importantly for the purpose of this {\it
Letter}---there is a gradual increase with radius in the age of the
youngest component of the dominant stellar population (from currently
active star formation  to 100 Myr, 0.8 Gyr, and 1.5 Gyr from the innermost
to the outermost field, respectively). This age gradient in the youngest
population is correlated with the HI column density as measured by
Staveley-Smith et al.\ (2003): the two innermost fields are located
$\simeq$ 0.7 Kpc at either side of $R_{H\alpha}$, LMC0512-6648 on the
local maximum of the azimutally averaged HI column density with $\simeq
1.63\times 10^{21} {\rm cm}^{-2}$ (close to the HI threshold for star
formation; Skillman 1987) and LMC0514-6503 where the azimutally averaged
HI column density is only $\simeq 5\times 10^{20} {\rm cm}^{-2}$. 
Finally, the two outermost fields are close to the HI radius considered by
Staveley-Smith et al.\  at the HI density of $10^{20} {\rm cm}^{-2}$. The
outermost field, LMC0513-6159, is approximately halfway to the tidal
radius (van der Marel et al. 2002). If the youngest stars in each field
were formed {\it in situ}, we are observing an outside-in quenching of the
star formation at recent times ($\simeq$ 1.5 Gyr) in the LMC, possibly
implying a decrease in size of the HI disk able to form stars.
Alternatively, the youngest stars may be migrating outwards from their
formation site with higher HI density (e.g.\ Ro\u{s}kar et al.\ 2008). In
fact, it is expected that both star formation sites and stars migrate
across the LMC disk due to tidal interactions with the Milky Way and the
SMC (e.g. Bekki \& Chiba 2005).

Taylor et al.\ (2005) offered two possible explanations for the fact that
the late-type galaxies in their sample  become redder outwards: a change
in the mean age of stellar population, or a change in the dust
characteristics. The present work shows that the youngest, bluest stars
are progressively missing outwards in the LMC disk. Therefore, a reddening
of the intrinsic integrated color is expected: for example, from the
synthetic population used in this paper, we measure (V-I) colors of 0.85,
0.97 or 1.03 if star formation continues to the present time, or is
truncated 0.8 or 1.5 Gyr ago, respectively. Other  irregular galaxies 
(and M33) are not close enough for such a detailed picture to be obtained,
but the evidence from shallower CMDs is that the situation is similar, so
that current star formation is concentrated in their central parts, and
the youngest population becomes gradually older outwards.  This is
observational evidence against the inside-out disk formation scenario in
the case of small spirals and irregular galaxies. 

Alternatively, it could be that the age distribution of the stars varies
with radius because of the interplay between the evolution of the
star-forming region (as a consequence of gas accretion and consumption)
and the outwards migration of the stars, as in the simulations by
Ro\u{s}kar et al.\ (2008; which however apply only to isolated, non-barred
galaxies). In this case, a position in between the two innermost
fields---which approximately coincides with the outermost extension of the
HI disk with density above the threshold for star formation---could be the
maximum radius reached by a growing star-forming zone and may  coincide
with the break radius observed in other more distant galaxies. In fact, a
break in the surface brightness profile at R$\simeq$240\arcmin\ is hinted
at in Figure 1\footnote{In that figure, the radius was calculated from the
center of the LMC bar; R=300\arcmin\ in that figure corresponds
approximately to R=240\arcmin\ from the LMC kinematic center} of Gallart
et al.\ (2004) and also in van der Marel (2001).







\acknowledgments

C.G. acknowledges interesting discussions with  A. Aparicio, M. Balcells,
K. Bekki, E. Bernard, S. Cassisi, V. Debattista, B. Elmegreen,  J.C.
Mu\~noz-Mateos, N. N\"oel, I. P\'erez, J. Read,  I. Trujillo, R. Zinn and
A. Zurita. The data presented in this paper was obtained as part of a
joint project between the University of Chile and Yale University funded by the
Fundaci\'on Andes. C.G. acknowledges partial support from the IAC and the
Spanish MEC (AYA2004-06343). This work has made use of the IAC-STAR
Synthetic CMD computation code. IAC-STAR is supported and maintained by
the computer division of the IAC.  {\it Facilities:} \facility{Blanco ()}

\clearpage


\end{document}